С.С. Федушко, асистент,
Ю.Р. Бекеш, студент
Національний університет «Львівська політехніка»
felomia@gmail.com

# Специфіка позиціонування послуг туристичної фірми у соціальних мережах

*У цій статті розглянуто методи формування клієнтської бази туристичної фірми за допомогою соціальних мереж. Позиціонування послуг туристичної фірми "Нова Європа" в мережі Інтернет здійснено методом залучення веб-користувачів соціальних мереж (VK.com та Facebook). Також використовуються методи застосування заходів з обслуговування та інтересів веб-користувачів.*

***Ключові слова:*** *соціальна мережа, туристична фірма, брендинг, клієнтська база, критерій ефективності, позиціонування.*

## Вступ

На світовому рівні в умовах глобалізації використання сучасних інформаційних технологій при наданні туристичних послуг є пріоритетним завданням. Оскільки немає чіткого позиціонування та узгодженості дій у промоції туристичних послуг, які дає позиціонування у соціальних мережах.

Просування в соціальних мережах – комплекс заходів по використанню соціальних медіа для просування компаній і вирішення різноманітних бізнес-завдань в мережі Інтернет [1].

Основне завдання просування бренду в соціальних мережах – популяризувати бренд та послугу в Інтернеті з допомогою проведення рекламної кампанії в соціальних мережах.

Остання тенденція зумовлює необхідність активізації туристичних компаній щодо їх інформаційного позиціонування у віртуальному просторі на основі використання брендів [2].

Особливо актуальні маркетингові технології, які спрямований на створення та управління бренду, для новостворених фірм. Згідно зі статистикою [3], щороку закривається майже 90% новостворених фірм.

Дослідженням вищенаведених задач займалися Водолазька С. [4], яка проаналізувала особливості застосування соцмереж як ефективного інноваційного способу просування видавництва і видавничої продукції на ринок. Також Фісенко Т. В. [5] розглядав основні принципи та засади просування спільнот соціальних інтернет-мереж.

Мороз М. [6] досліджував напрями застосування сайтів соціальних мереж в просуванні продукту, а також процеси, які відбуваються під час підготовки і реалізації головної проблеми компанії – формування сприйняття клієнтом торгової марки, залучення нових клієнтів. Теоретичні принципи формування та подальшого розвитку груп і співтовариств у соціальних мережах Інтернету, які покращують свою позицію в глобальному інформаційному середовищі розглядав Колакзик Е. [7].

Зуб Т. А. та Зозульов О. В. [8] досліджували сутність засобів телекомунікації та специфіка їх застосування для просування послуг. Особливу увагу приділяли Інтернет-комунікаціям, виділили переваги від просування через мережу Інтернет. Також, проводили порівняння традиційних ЗМІ з мережею Інтернет, при цьому розглядали комунікаційні моделі.

У своїх наукових дослідженнях Кетов Н. П. [9] виклав особливості реклами в Інтернеті, розкрив її специфіку в соціальних мережах. Обґрунтував фактори просування товарних брендів, охарактеризував інструменти зростання ефекту при креативному підході до організації рекламних кампаній в соціальних мережах.

К. Тукер [10] досліджувала у своїй праці як інтернет-користувачі сприймають контроль над їхньою особистою інформацією і скільки є шансів на те, що вони клікатимуть на інтернет-рекламу.

## Специфіка позиціонування турфірми

Туристичні фірми орієнтуються на широку аудиторію споживачів для того, щоб забезпечувати постійний попит на туристичні послуги, тому доцільно створювати їхні профілі у таких соціальних мережах як VK.com та Facebook. Проте, наявність профілю туристичної фірми в певній соціальній мережі є лише першим кроком до ефективного використання такого інструменту інтернет-технологій.

Профіль у соціальній мережі можна вважати певним прототипом веб-сайту туристичної фірми, але тут оновлення та актуалізація інформації здійснюється лише одним виконавцем.

*Основні переваги просування туристичної фірми в соціальних мережах:*

− Індивідуальне спрямування (демографічні показники, інтереси, сімейний статус, регіон). Є можливість максимально точного визначення потенційних клієнтів.

− Масштаби охоплення. Понад 66% усього населення України, а отже і потенційних клієнтів, є користувачами соціальних мереж.

− Відсутність мінімального бюджету. Всі пропозиції розробляються індивідуально і мають свої неповторні «фішки».

− Фактор вірусного маркетингу. Люди самі показують рекламу фірми [4].

− Таргетинг – демонстрація реклами чітко визначеній аудиторії.

− Трекінг – можливість аналізу поведінки відвідувачів сайту і вдосконалення сайту, продукту і особливостей проведення маркетингової діяльності відповідно до висновків такого аналізу.

− Доступність (за принципом 24 години на добу, 7 днів на тиждень) і гнучкість (почати, редагувати і перервати рекламну кампанію можна миттєво).

− Забезпечення доступної інформації про фірму або послуги для значної за чисельністю кількості людей, у тому числі – для географічно віддалених об'єктів.

− Інтерактивність – споживач взаємодіє з продавцем і з продуктом, вивчає його.

− Можливість розміщення великої кількості інформації (графіки, звуку, відео тощо).

− Оперативність поширення і отримання інформації.

− Швидкий зворотний зв'язок з цільовою аудиторією, рівного якому немає в жодній з традиційних форм реклами.

− Порівняно низька вартість.

− Значний візуальний канал впливу на людину, можливість більш пильної уваги користувача перед комп'ютером, концентрації на деталях.

− Створення віртуальних груп за інтересами.

− Автоматичне отримання рекламодавцем інформації, яка у традиційній рекламі потребує дорогих досліджень [12].

*Недоліки просування послуг туристичної фірми у соціальних мережах:*

− Збільшення недовіри інтернет-рекламі через бурхливий розвиток фіктивних пропозицій у перші роки існування такої послуги.

− Приховування реальної статистики власниками веб-сайтів або банерних мереж, тобто штучно підвищують статистичну кількість відвідувань інтернет-ресурсів.

− Складність знайдення потрібного сайту з потрібною рекламою [13].

− Залучення великої кількості користувачів розцінюється як спам.

− Блокування сторінки адміністратором на підставі скарг користувачів.

Враховуючи всі переваги і недоліки таку концепцію позиціонування бренду туристичної фірми у соціальних мережах застосовано до створення позитивного інформаційного образу туристичної фірми "Нова Європа".

Створення позитивного інформаційного образу туристичної фірми "Нова Європа" полягає у реалізації комплексу методів популяризації послуг фірми серед користувачів соціальних мереж. Зважаючи на специфіку надання послуг організації для певної соціальної мережі потрібно індивідуально розробляти такий комплекс методів позиціонування. Зокрема, для створення позитивного інформаційного образу турфірми "Нова Європа" у соціальній мережі "Vk.com" необхідно створити тематичну групу і користувача з метою залучення клієнтів до цієї групи. Комплекс включає такі завдання:

− Коректне створення зручної для користування групи та користувача фірми "Нова Європа";

− Наповнення атрактивним контентом.

− Розсилання листів-запрошень з метою залучення потенційних клієнтів фірми.

− Контекстна та банерна реклама.

− Щоденне модерування стіни групи.

− Залучення потенційних клієнтів, розміщуючи посилання група, сайтах та веб-спільнотах.

− Постійне оновлення інформації, розміщення цікавого та актуального контенту та інтерактивна комунікація з учасниками.

− Слідкування за статистикою групи та внесення коректив до управління групою.

− Розміщення посилань на сайті фірми "Нова Європа" [14].

Використовуючи вищезазначені методи, кількість учасників групи "Нова Європа" активно зростає. Відповідно до статистики [15], за період її існування тематичної групи кількість учасників становить 225 учасника у Vk.com (рис. 1) і щоденно збільшується приблизно на 3-4 учасники.

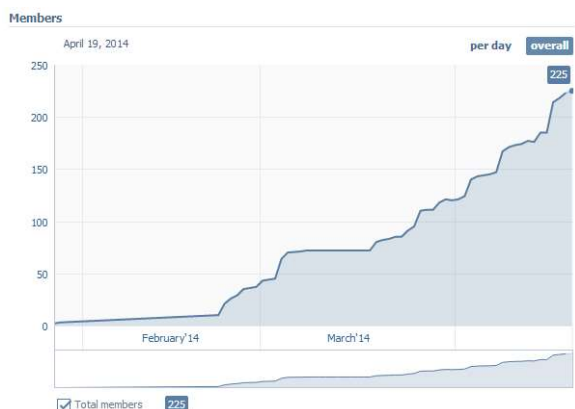

Рисунок 1 – Динаміка збільшення учасників групи туристичної фірми "Нова Європа" у Vk.com

Відмінністю Facebook від Vk.com є те, що тут сторінки можуть переглядати навіть незареєстровані користувачі, а для того, щоб відслідковувати оновлення, достатньо натиснути "мені подобається".

Власна група або сторінка компанії, в цьому випадку "Нова Європа" дозволяє відразу ж повідомляти численним користувачам Facebook про всі події, що пов'язані з компанією. Крім того, в Facebook можна проводити різні рекламні заходи: акції, конкурси. [16].

Для популяризації групи у Facebook важливу роль відіграє спілкування в групах, оскільки це надає більшої уваги профілю у якому розміщене посилання на головну сторінку "Нової Європи".

У Facebook [17] також спостерігається висока активність учасників (рис.2).

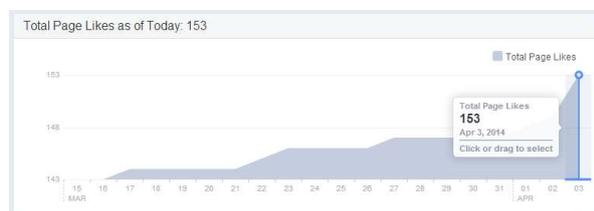

Рисунок 2 – Динаміка збільшення учасників групи у Facebook

Для підтримки активності учасників групи та отримання інформації про переваги та недоліки функціонування фірми проводяться опитування та обговорення. Обмеження можливості запрошення певної категорії користувачів до групи істотно звужує коло нових учасників.

До переваг застосування комплексу створення позитивного інформаційного образу туристичної фірми "Нова Європа" можна віднести те, що соціальної мережа Vk.com дає можливість фільтрації своїх користувачів при пошуку, тобто вибирати саме цільову аудиторію користувачів, відповідно до соціально-демографічного портрету потенційних клієнтів фірми.

Щодо Facebook, то між групами або сторінками з точки зору просування сайтів є багато відмінностей. Вони полягають у різному призначенні цих видів представництва компаній в одній із найпопулярніших соціальних мереж. Наприклад, метою створення груп у Facebook є спілкування між учасниками, обговорення різних тем. Сторінки в Facebook, призначені для того, щоб представляти компанії у цій соціальній мережі. Це може бути як представництво компанії взагальному, так і її конкретних послуг.

У Facebook можна встановлювати додатки, які будуть привертати увагу користувачів, а також є можливість переглядати статистику відвідувань. Найефективнішим способом просування сайтів у Facebook це створення сторінок.

Для більшого заохочення користувачів відвідувати сторінку в Facebook, можна розробити для неї спеціальний дизайн та розмістити цікавий і корисний контент.

Найефективнішим способом просування сайтів у Facebook це створення сторінок. Для більшого заохочення користувачів відвідувати сторінку в Facebook, можна розробити для неї спеціальний дизайн та розмістити цікавий і корисний контент.

Експертами здійснено ґрунтовний порівняльний аналіз критерії оцінювання ефективності роботи з соціальною мережею в галузі позиціонування бренду туристичної фірми у глобальному інформаційному просторі, а саме, у соціальних мережах. Результати цього аналізу представлені у Таблиці 1.

*Таблиця 1*
Порівняльна характеристика доступних критеріїв ефективності позиціонування бренду туристичної фірми у Vk.com та Facebook

| Позначення | Критерії ефективності позиціонування | Значення | |
|---|---|---|---|
| | | Vk.com | Facebook |
| Критерій 1 | додавання фото/відео | 8 | 9 |
| Критерій 2 | перегляд медіа матеріалів | 5 | 9 |
| Критерій 3 | створення опитувань | 7 | 9 |
| Критерій 4 | участь в обговореннях | 6 | 8 |
| Критерій 5 | меню групи | 10 | 10 |
| Критерій 6 | розміщення контекстної та банерної реклами | 7 | 8 |
| Критерій 7 | щоденне модерування стіни групи | 7 | 10 |
| Критерій 8 | розміщення посилань у інших група, сайтах та веб-спільнотах | 7 | 8 |
| Критерій 9 | розміщення цікавого та актуального контенту | 10 | 10 |
| Критерій 10 | інтерактивна комунікація з учасниками групи | 10 | 10 |
| Критерій 11 | внесення корективи до управління групою | 8 | 7 |
| Критерій 12 | додавання нових учасників | 7 | 9 |
| Критерій 13 | відправлення особистих повідомлень учасникам групи | 10 | 10 |
| Критерій 14 | публікація новин | 9 | 10 |
| Критерій 15 | завантаження файлів | 7 | 9 |
| Критерій 16 | створення заходів | 10 | 10 |
| Критерій 17 | затрата часу адміністратора | 8 | 8 |
| Критерій 18 | швидкість поширення інформації | 7 | 8 |

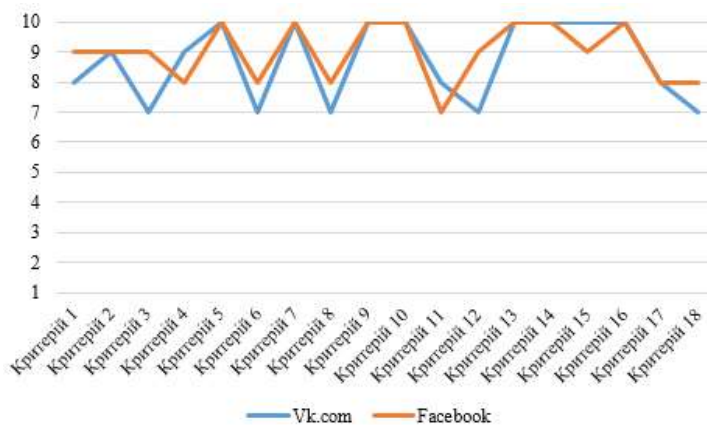

Рисунок 3 – Графік критерії оцінювання ефективності роботи з соцмережами у Vk.com та Facebook

Для оцінки зручності роботи у соціальної мережі було введено критерії оцінювання ефективності роботи з соціальною мережею.

Причому, $\text{Criterion} \in [0;10]$. На основі цих критеріїв побудовано комплексний показник, як усереднене значення цих критеріїв.

Обчислення цього показника здійснюється за допомогою формули 1 для соціальні мережі Facebook, а для соціальні мережі VK.com за допомогою формули 2.

$$P_{\text{efficiency}}(Fb) = \frac{\sum_{j=1}^{N} \text{Criterion}(Fb)_j}{N}, \quad (1)$$

де $\sum_{j=1}^{N} \text{Criterion}(Fb)_j$ - сума критерії ефективності позиціонування бренду у соціальній мережі Facebook; $N$ - кількість цих критеріїв.

$$P_{\text{efficiency}}(Vk) = \frac{\sum_{j=1}^{N} \text{Criterion}(Vk)_j}{N}, \quad (2)$$

де $\sum_{j=1}^{N} \text{Criterion}(Vk)_j$ - сума критерії ефективності позиціонування бренду у соціальній мережі VK.com; $N$ - кількість цих критеріїв. На основі цих даних побудовано графік (див. рис. 3).

**Соціально-демографічний портрет учасників спільнот турфірми у соцмережах.**

Важливим етапом у позиціонуванні бренду турфірми є створення соціально-демографічного портрету (соціально-демографічний портрет учасника веб-спільнот – набір верифікованих базових соціально-демографічних характеристик учасника віртуальної спільноти, який формується на основі комп'ютерно-лінгвістичного аналізу інформаційного сліду учасника веб-спільноти [18, 19]) потенційних клієнтів та перевірки достовірності соціально-демографічних характеристик (соціально-демографічні характеристики – сукупність соціальних критеріїв оцінки та важливих параметрів особистості. Наприклад, вік, стать тощо. [20, 21]) з метою створення клієнтської бази.

Особливо, це завдання є актуальним у глобальній мережі Інтернет, адже кількість користувачів групи не є пріоритетним завдання, а якісний підбір користувачів – потенційних клієнтів послуг фірми.

У цьому дослідженні здійснено аналіз соціально-демографічного портрету учасника тематичних груп тур фірми "Нова Європа" у соціальних спільнотах. Візуально соціально-демогарафічний портрет (вік та стать) учасників групи у соціальної мережі Facebook представлено на рис. 4 та соціальної мережі Vk.com – на рис. 5.

Як бачимо з рис. 4 та 5 жінки більш часті відвідувачі та учасники спільноти туристичної фірми "Нова Європа", як у Facebook, так і у Vk.com. У соціальної мережі Vk.com учасники тематичної групи здебільшого належать до вікових груп до 18 років та 21-24 років. В свою чергу учасники соціальної мережі Facebook є більш зрілою аудиторією, які здатні оплатити послуги туристичної фірми. Проте, множина учасників молодшої вікової групи теж є потенційними клієнтами, оскільки є ймовірність спонсорської допомоги батьків.

Отож, зважаючи на все вище сказане, можна зробити висновок, що доцільно позиціонуванні бренду тур фірми у двох цих спільнотах, щоб охопити всю множину потенційних клієнтів.

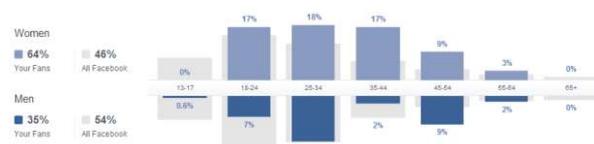

Рисунок 4 – Соціально-демогарафічний портрет учасників групи у соцмережі Facebook

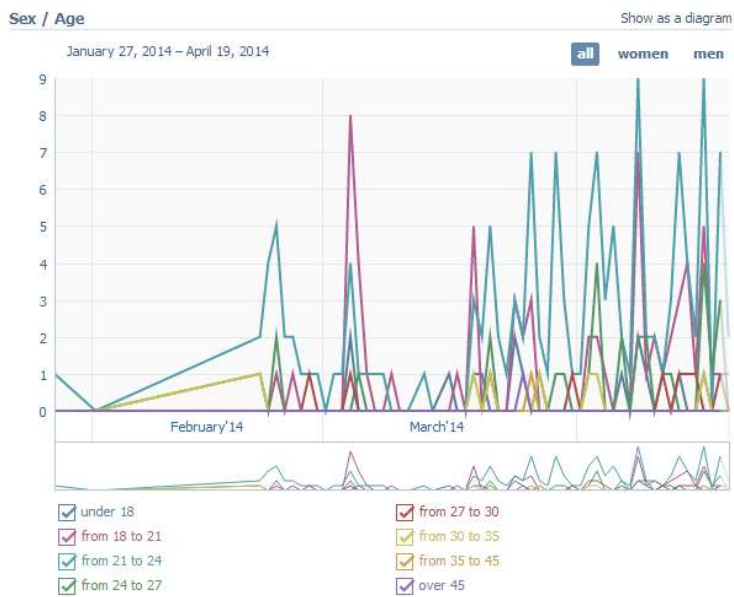

Рисунок 5 – Соціально-демогарафічний портрет учасників групи у соцмережі Vk.com

## Висновок

Сьогодні використання інформаційно-комунікаційних технологій є ефективним засобом для створення позитивного інформаційного образу, залучення та інформування потенційних клієнтів туристичної фірми "Нова Європа". У цій статті досліджено специфіку інформаційного позиціонування туристичної фірми у віртуальному просторі на основі використання соціальних мереж Vk.com та Facebook та здійснено порівняльний аналіз доступних методів позиціонування бренду туристичної фірми у цих соціальних мережах. Введено критерії оцінювання ефективності роботи з соціальною мережею, на основі яких побудовано комплексний показник. Також, проаналізовано соціально-демографічного портрету учасника тематичних груп тур фірми "Нова Європа" у соціальних спільнотах.

**С.С. ФЕДУШКО, Ю.Р. БЕКЕШ**
Национальный университет «Львовская политехника»

**СПЕЦИФИКА ПОЗИЦИОНИРОВАНИЯ УСЛУГ ТУРИСТИЧЕСКОЙ ФИРМЫ В СОЦИАЛЬНЫХ СЕТЯХ**

В этой статье рассмотрены методы формирования клиентской базы туристической фирмы с помощью социальных сетей. Позиционирование услуг туристической фирмы "Новая Европа" в сети Интернет осуществлено методом привлечения веб-пользователей социальных сетей (VK.com и Facebook). Также используются методы применения мер по обслуживанию и интересов веб-пользователей.

*Ключевые слова: социальная сеть, туристическая фирма, брендинг, клиентская база, критерий эффективности, позиционирование.*

**S.S. FEDUSHKO, Yu.R. BEKESH**
Lviv Polytechnic National University

**POSITIONING SERVICES OF A TRAVEL AGENCY IN SOCIAL NETWORKS**

In this paper the methods of forming a travel company customer base by means of social networks are observed. These methods are made to involve web-users of the social networks (VK.com and Facebook) for positioning of the service of the travel agency "New Europe" on the Internet. The methods of applying the maintenance activities and interests of web-users are also used. So, the main method of information exchanging in modern network society is on-line social networks. The rapid development and improvement of such information and communication technologies is a key factor in the positioning of the travel agency brand in the global information space. The absence of time and space restrictions and the speed of spreading of the information among an aim audience of social networks create all the conditions for effective popularization of the travel agency "New Europe" and its service in the Internet.

***Keywords:** social networking, travel agency, branding, customer base, efficiency criteria, positioning.*